\documentclass[draft]{agujournal2019}
\usepackage{url} %this package should fix any errors with URLs in refs.
\usepackage{lineno}
\usepackage{xcolor}
\usepackage[inline]{trackchanges} %for better track changes. finalnew option will compile document with changes incorporated.
\usepackage{soul}
\draftfalse

\journalname{Earth and Space Science}

\begin{document}

%% ------------------------------------------------------------------------ %%
%  Title

\title{Atmospheric Temperature Effect in secondary cosmic rays observed with a two square meter ground-based detector}

%% ------------------------------------------------------------------------ %%

%  AUTHORS AND AFFILIATIONS

\authors{Irma Ri\'adigos\affil{1}, Dami\'an Garc\'ia-Castro\affil{2}, Diego Gonz\'alez-D\'iaz\affil{2}, and Vicente P\'erez-Mu\~nuzuri\affil{1}}

\affiliation{1}{CRETUS Institute, Group of Nonlinear Physics, Faculty of Physics, University of Santiago de Compostela, Spain}
\affiliation{2}{Instituto Galego de F\'isica de Altas Enerx\'ias, Faculty of Physics, University of Santiago de Compostela, Spain}

%% Corresponding Author:
% Corresponding author mailing address and e-mail address:

\correspondingauthor{Irma Ri\'adigos}{irma.riadigos@usc.es}

%% Keypoints, final entry on title page.

%  List up to three key points (at least one is required)
%  Key Points summarize the main points and conclusions of the article
%  Each must be 140 characters or fewer with no special characters or punctuation and must be complete sentences

\begin{keypoints}
\item Atmospheric temperature effect isolated with timing resistive plate chambers for the first time
\item Extraction of temperature coefficients versus atmosphere depth through principal component regression and comparison with theoretical expectation
\item Seasonal and occasional variations for the cosmic ray flux observed, showing consistency with earlier Global Muon Detector Network results
\end{keypoints}

%% ------------------------------------------------------------------------ %%
%
%  ABSTRACT and PLAIN LANGUAGE SUMMARY
%
%% \begin{abstract} starts the second page

\begin{abstract}
\justify
A high resolution 2 m$^2$ tracking detector, based on timing Resistive Plate Chamber (tRPC) cells, has been installed at the Faculty of Physics of the University of Santiago de Compostela (Spain) in order to improve our understanding of the cosmic rays arriving at the Earth's surface. Following a short commisioning of the detector, a study of the atmospheric temperature effect of the secondary cosmic ray component was carried out. 
A method based on Principal Component Analysis (PCA) has been implemented in order to obtain the distribution of temperature coefficients, $W_T(h)$, using as input the measured rate of nearly vertical cosmic ray tracks, showing good agreement with the theoretical expectation. The method succesfully removes the correlation present between the different atmospheric layers, that would be dominant otherwise. We briefly describe the initial calibration and pressure correction procedures, essential to isolate the temperature effect. Overall, the measured cosmic ray rate displays the expected anticorrelation with the effective atmospheric temperature, through the coefficient $\alpha_T= -0.279 \pm 0.051 $ \%/K. Rates follow the seasonal variations, and unusual short-term events are clearly identified too.
\end{abstract}

%% ------------------------------------------------------------------------ %%
%
%  TEXT
%
%% ------------------------------------------------------------------------ %%

\justify

\section{Introduction}

Cosmic rays (c.r.) are messengers from outer space that provide valuable information for different research areas, such as space weather, high energy physics and cosmology \cite{pierre,grapes}. Specifically, ground-based instruments give us the chance to study the products of the interactions between primary cosmic rays and the nuclei in the atmosphere. These products consist of secondary particles that traverse the atmosphere carrying information about its inner structure, as in a radiography, providing information about its properties. Secondary muons, in particular, are affected by atmospheric pressure and temperature. These induce local modifications of the atmospheric density and its depth, thereby changing the balance between particle production, absorption and decay, affecting the muon rates at ground \cite{dorman}. Since modern muon detectors are mostly committed to the study of solar activity and other astrophysical phenomena, such effects are regularly removed with simple techniques, as part of a calibration procedure \cite{mendonca_simple}. Our work is concerned with a deeper comprehension of such atmospheric effects. 

The rates of c.r. measured at the Earth's surface vary according to changes in several atmospheric conditions, chiefly the pressure at ground level and the temperature profile of the atmosphere. The pressure (or barometric) effect arises mainly due to the absorption of the radiation in the atmosphere through energy loss. The more the pressure, the more air mass to traverse. Therefore, secondary c.r. particles are more efficiently absorbed and, consequently, the measured rates are modulated in anticorrelation with the ground-level pressure. In detail, the mass increase is associated with a larger height of the air column, that modifies the decay probabilities of the particles involved, implying a non-negligible (but sub-dominant) contribution to the main negative correlation \cite{sagisaka}.

In the absence of variations in the ground-level pressure, the modification of the local density of the air column as a result of temperature variations modifies c.r. rates as well. This so-called `temperature effect' is the result of two contributions: the interaction (positive correlation) and decay (negative correlation) of secondary particles through the atmosphere. Muons ($\mu^{+/-}$), that are the dominant c.r. particles at ground level, originate mainly from the decay of charged pions and kaons ($\pi^{+/-}, \textnormal{K}^{+/-}$). Since an increase in temperature reduces the atmospheric density, causing a reduction in the interaction probability, more pions and kaons will decay and produce more muons. Muons, unlike their parents, are highly penetrating, and small changes in their interaction probability leave their behaviour largely unaffected. As a result, the surface rates increase with increasing temperature, creating a positive effect. The negative effect is related to the muon decay itself. Because the atmosphere also expands when the temperature increases, muons have to travel further to reach the surface before they decay. As a result, the surface rates decrease. Importantly, in case of high-energy particles, the positive effect dominates because of time dilation in the muon reference frame. This relativistic effect causes energetic muons to effectively live longer, so their decay probability is reduced. For low energies, on the other hand, the dominant effect is negative. A practical way to `increase' the particle energy is to measure the c.r. rates underground, thus allowing to study the temperature effect as a function of the rock overburden (e.g., \cite{doublechooz}).

The temperature effect can be nearly one order of magnitude smaller than its pressure counterpart, enforcing a much better control on systematic effects of instrumental origin and high statistics (detector size) in the first place. A precise estimate of the atmospheric temperature effect involves an implementation of the `integral method', which requires, on the one hand, knowing the temperature profiles above the detector and, on the other hand, knowing the distribution of temperature coefficients ($W_T$), the latter accessible by theoretical means \cite{dmitrievacoef}. Given that the detector used in this work was conceived to potentially make use of both the soft (electron) and hard (muon) cosmic ray component, the temperature coefficients (calculated for muons) are not known a priori. As discussed afterwards, the effect of the material overburden can not be completely neglected, nor easily characterized, either. 

The main purpose of this work is to experimentally obtain the temperature coefficients for secondary cosmic rays at ground, resorting to a technology not used previously in these kind of studies. We will show how, despite the much higher dark rates customary of gaseous detectors as compared to plastic scintillators, two $\sim 2$ m$^2$ planes of timing Resistive Plate Chambers (tRPCs) operating at ground level are sufficient to isolate the temperature effect, a fact that results largely from the superb timing characteristics of the device.

\section{The Muon Telescope (MT)}

The tRPC technology was introduced to particle physics back in 2000 as a byproduct of the R\&D program of the ALICE experiment at the Large Hadron Collider \cite{alice}. Indeed, tRPCs have been adopted already for the study of high energy cosmic rays by the EEE collaboration \cite{EEE}, but no study on the temperature effect has been reported by the collaboration yet. tRPCs represent a family of non-proportional gaseous detectors, generally characterized by the use of thin sub-mm gas gaps operated in `fast' well-quenched gas mixtures at very high electric fields (up to $\sim$ 150 kV/cm). Stability of operation requires the use of insulating materials with high surface quality, something conventionally achieved through soda-lime glass. tRPCs can make an optimal use of the multi-gap technique \cite{Multigap}, that allows for the systematic stacking of several gas gaps, in order to achieve time resolutions down to 20 ps in special configurations \cite{Williams24}. In fact, tRPCs have recently demonstrated the capability of reaching 60 ps on $2\times2$ m$^2$ areas, with a modest number of electronic readout channels around 160, and a position resolution at the cm-scale \cite{Watanabe}.

\begin{figure}
\includegraphics[width=\textwidth]{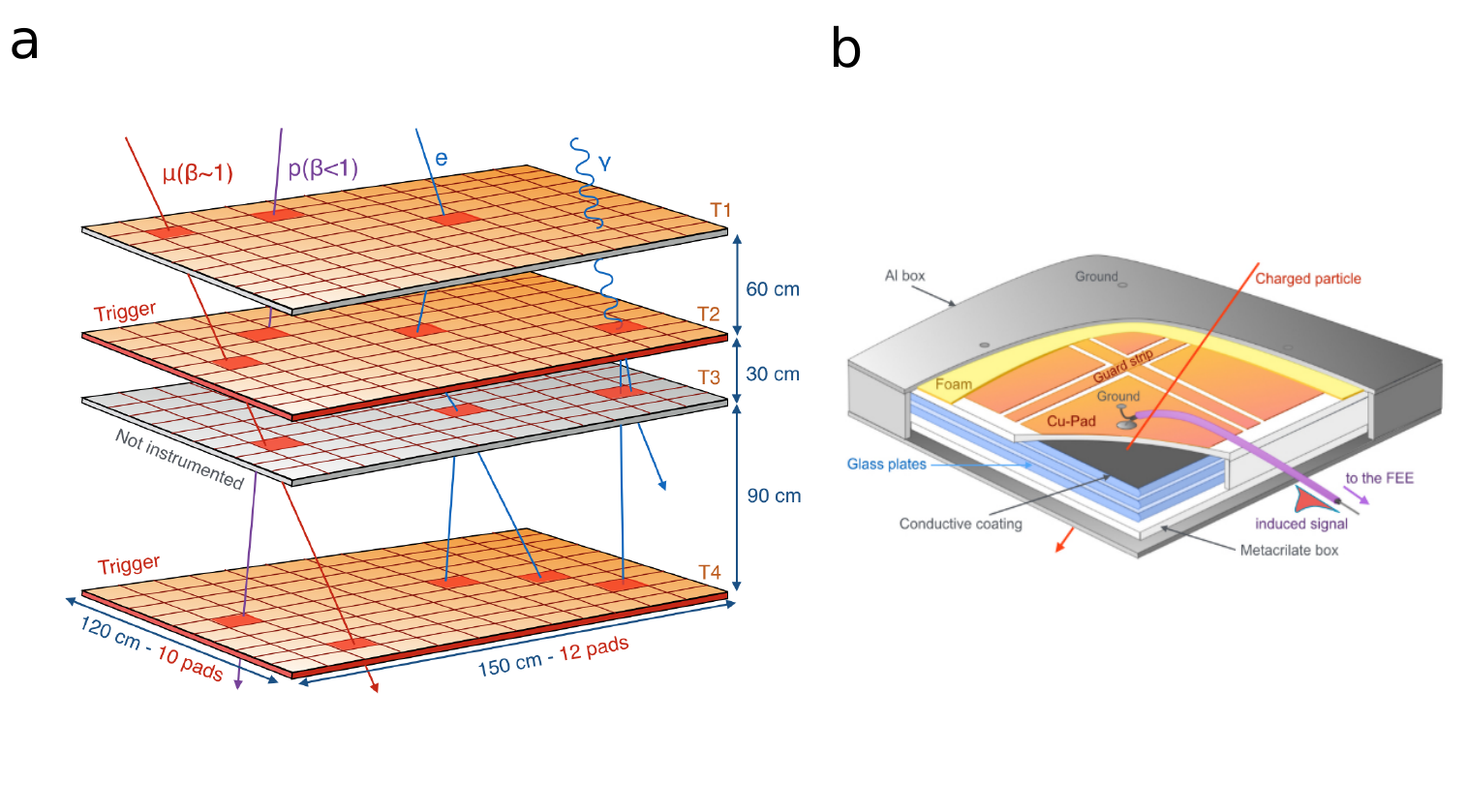}
\caption{(a) Drawing of the muon telescope at Santiago de Compostela, showing the tRPC layout with some illustrative examples of different particle interactions. In the present analysis only the T2 and T4 RPC planes are used. (b) Inner view of the detector.}
\label{fig:tragal_draw}
\end{figure}

At the University of Santiago de Compostela (Spain), a medium-size tRPC detector (1.2 $\times$ 1.5 m$^2$) with a space and time resolution of $\sigma_{x,y}\sim3$ cm and $\sigma_t \sim 300$ ps, respectively, has been installed circa 2014 \cite{tragaldabas}. It is named TRAGALDABAS (TRAsGo for the AnaLysis of the nuclear matter Decay, the Atmosphere, the earth B-Field And the Solar activity), and it has been designed and built at LIP-workshops (e.g., \cite{LuisMarta}). The angular resolution with its present vertical layout is 2-3$^\circ$ and the maximum zenith angle of the accepted tracks is close to 50$^\circ$ (Figure \ref{fig:tragal_draw}a). Unlike other cosmic ray detectors, TRAGALDABAS has a relatively small active area of 1.8 m$^2$. For comparison, the MuSTAnG detector \cite{mustang} had a surface of 4 m$^2$ and all four telescopes of the Global Muon Detector Network (GMDN) extend over around 15-30 m$^2$ \cite{GMDN}. 

The MT consists of four RPC planes with a total height of 1.8 m (Figure \ref{fig:tragal_draw}a). Each plane's inner design is based on three plates of 2 mm glass with a 1 mm gas gap interleaved, placed inside a gas-tight acrylic box. Tetrafluoroethane, a type of freon (R134a, CF$_3$CH$_2$F), is used as the active medium, at a very low flow, just sufficient to keep the detector efficiency constant over time. The external sides of the outer gas plates are covered with a semi-conductive coating to which a $\pm$ 5600 V high voltage is applied. Electrical pick-up signals, stemming from the avalanches produced in the gas upon the passage of a charged particle, are induced in some of the 120 copper pads placed outside of the acrylic box (Figure \ref{fig:tragal_draw}b). Those signals are processed with fast $\sim 1$ GHz BW electronics \cite{DANIFEE} and, if above an adjustable threshold, a digital LVDS signal is produced, marking the passage of the particle (we will refer to the associated pad and plane as `fired'). A flexible trigger condition can be formed for any number of fired planes and pad multiplicity per plane, a digital signal formed, correspondingly, and sent to the acquisition in order to store the c.r. candidate. The telescope is placed at $\sim$260 m above sea level, $42^\circ$52'N $8^\circ$33'W, at a geomagnetic rigidity cutoff of $\sim$5.5 GV and in the first floor of a two-story building. It is running since 2015 with a room temperature stable at 20 $\pm$ 1$^\circ$C.

It is important to note that, except in special configurations (e.g. \cite{neutrons1}, \cite{neutrons2}), RPC detectors have intrinsically a very low detection efficiency for neutrons below 10~MeV, not exceeding 0.1\%. Moreover, neither the products from neutron interactions nor electrons from neutron decay at these energies can traverse a second detection plane, as required in present analysis. Hence, if assuming a typical albedo neutron flux of $\sim$1 kHz/cm$^2$ \cite{neutrons3}, the detector can be effectively considered as neutron-blind, for the purposes of present analysis.

\section{Input data and processing}

We report here data from the commissioning phase and early physics run (from October 2015 to January 2017), where only two detector planes, stacked over a height of 120 cm, were used (T2 and T4 in Figure \ref{fig:tragal_draw}a). A trigger condition was defined as `at least one fired pad per plane, in time coincidence'. During data analysis, a standard equalization is performed in an automatic way, aimed at the correction of the channel-by-channel variations in the time offsets and signal amplification (along the lines of \cite{Korna}). Provided both charge and time information are stored for each pad, noise signals (displaying zero-charge) can be removed in the next processing step. Finally, `particle tracks' are formed by combinatorialy matching the fired pads in both planes with a velocity compatible with the speed of light, within a $3$-$\sigma_t$ interval, being $\sigma_t$ the time resolution of the detector. This produces the final data sample ready for physics analysis, where any instrumental effects should be greatly minimized. We use in this work a data sub-sample, corresponding to events with a single track (multiplicity $M=1$), and a zenith angle $\theta$ lower than 13$^\circ$. The former condition means that only cases with one fired pad per plane have been considered.

\begin{figure}[h!!!]
    \centering
    \includegraphics[scale=0.55]{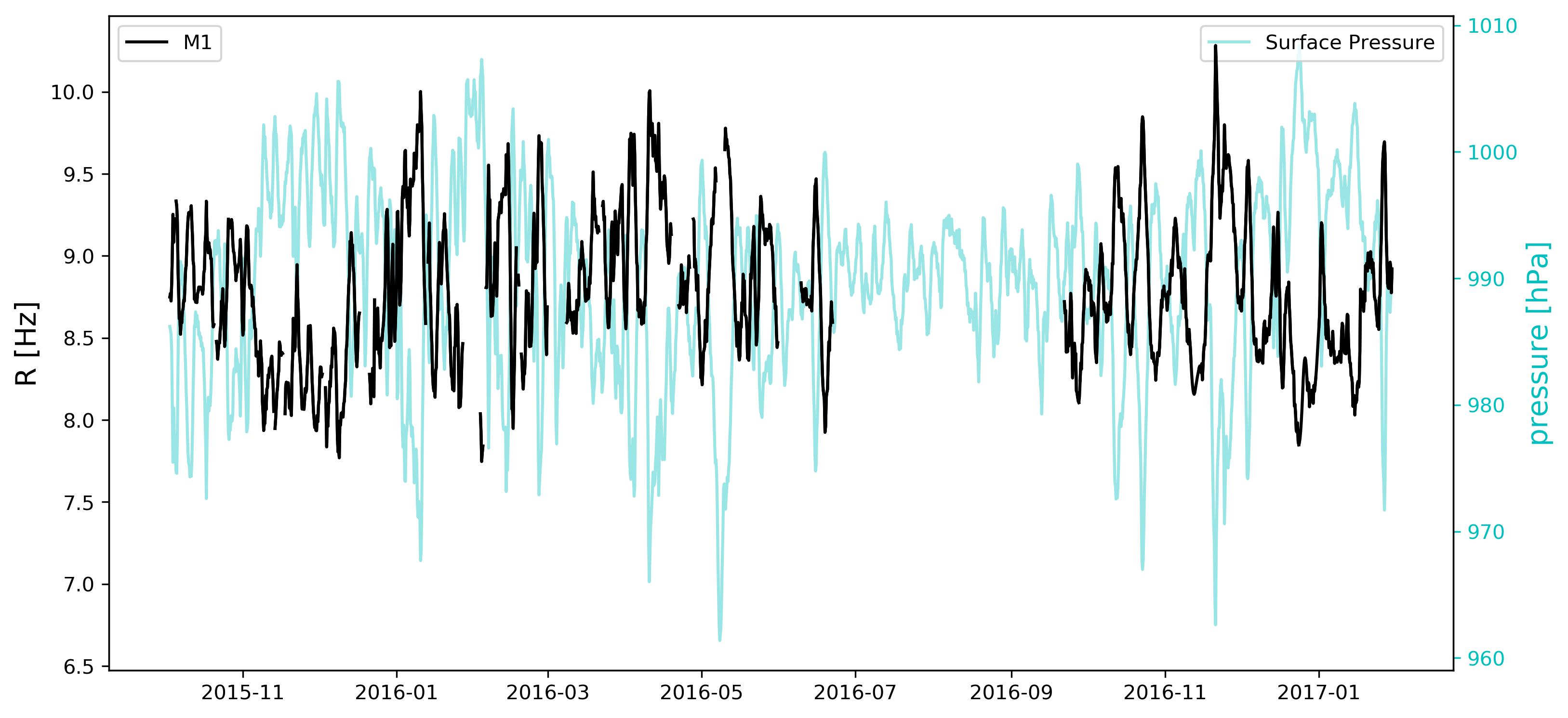}
    \caption{Cosmic ray rate for single vertical tracks as observed with the muon telescope (black curve) and ground-level pressure (blue curve).}
    \label{fig:data_raw}
\end{figure}

The complete data taking period, displayed in Figure \ref{fig:data_raw}, may be conveniently divided in three phases:

\begin{itemize}
\item From October 2015 to June 2016. During this first period the detector was run semi-autonomously (several interventions were needed) and problems related to faulty front-end electronics and high voltage instabilities were observed. 
\item From June 2016 to October 2016. Maintenance work was carried out, the faulty electronics modules were replaced, and an on-line monitor developed.
\item From October 2016 to January 2017. The detector run in stable conditions, in a fully autonomous way.
\end{itemize}

\begin{figure}[h!!!]
\begin{center}
\includegraphics[scale=0.5]{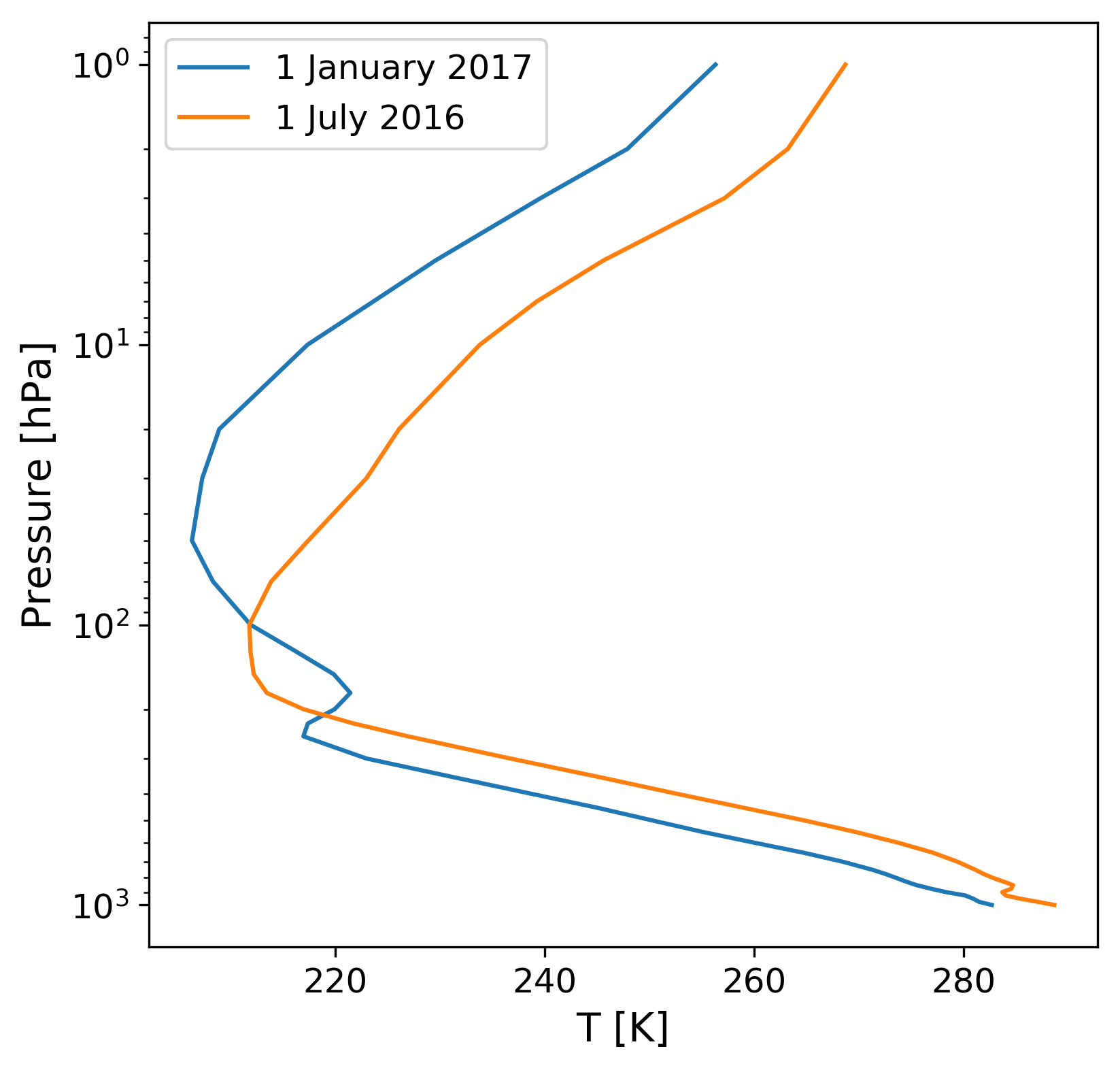}
\caption{Examples of ERA-INTERIM atmospheric temperature profiles for Santiago de Compostela, summer (1 July 2016) and winter (1 January 2017).}
\label{fig:temp_prof}
\end{center}
\end{figure}

Concerning the atmospheric variables, the vertical temperature profiles were retrieved from the European Centre for Medium-Range Weather Forecast (ECMWF) reanalysis, ERA-Interim \cite{erainterim}, for the 1979-2017 period, at 37 isobaric levels (1000, 975, 950, 925, 900, 875, 850, 825, 800, 775, 750, 700, 650, 600, 550, 500, 450, 400, 350, 300, 250, 225, 200, 175, 150, 125, 100, 70, 50, 30, 20, 10, 7, 5, 3, 2, 1 hPa),  with a horizontal spatial resolution of $0.125^\circ $ and a temporal resolution of 6 h. Two exemplary profiles for summer and winter are shown in Fig. \ref{fig:temp_prof}.

\section{Analysis of Atmospheric Effects}
As mentioned before, several methods can be used to take into account the temperature effect of secondary cosmic ray particles in the atmosphere \cite{blackett,berkova,dorman,duperier,sagisaka}. The integral method has been shown to be one of the most precise in most of the muon telescopes \cite{mendonca,dmitrievamethods} but it requires knowing the distribution of the temperature coefficients in the atmosphere, $W_T (h)$. These can be theoretically calculated for different threshold energies, zenith and azimuth angles of incidence \cite{dmitrievacoef}, or extracted from c.r. data \cite{yanchuk_pca}. In this work we compare both approaches.

\begin{figure} [h!!!]
\begin{center}
\includegraphics[width=0.5\textwidth]{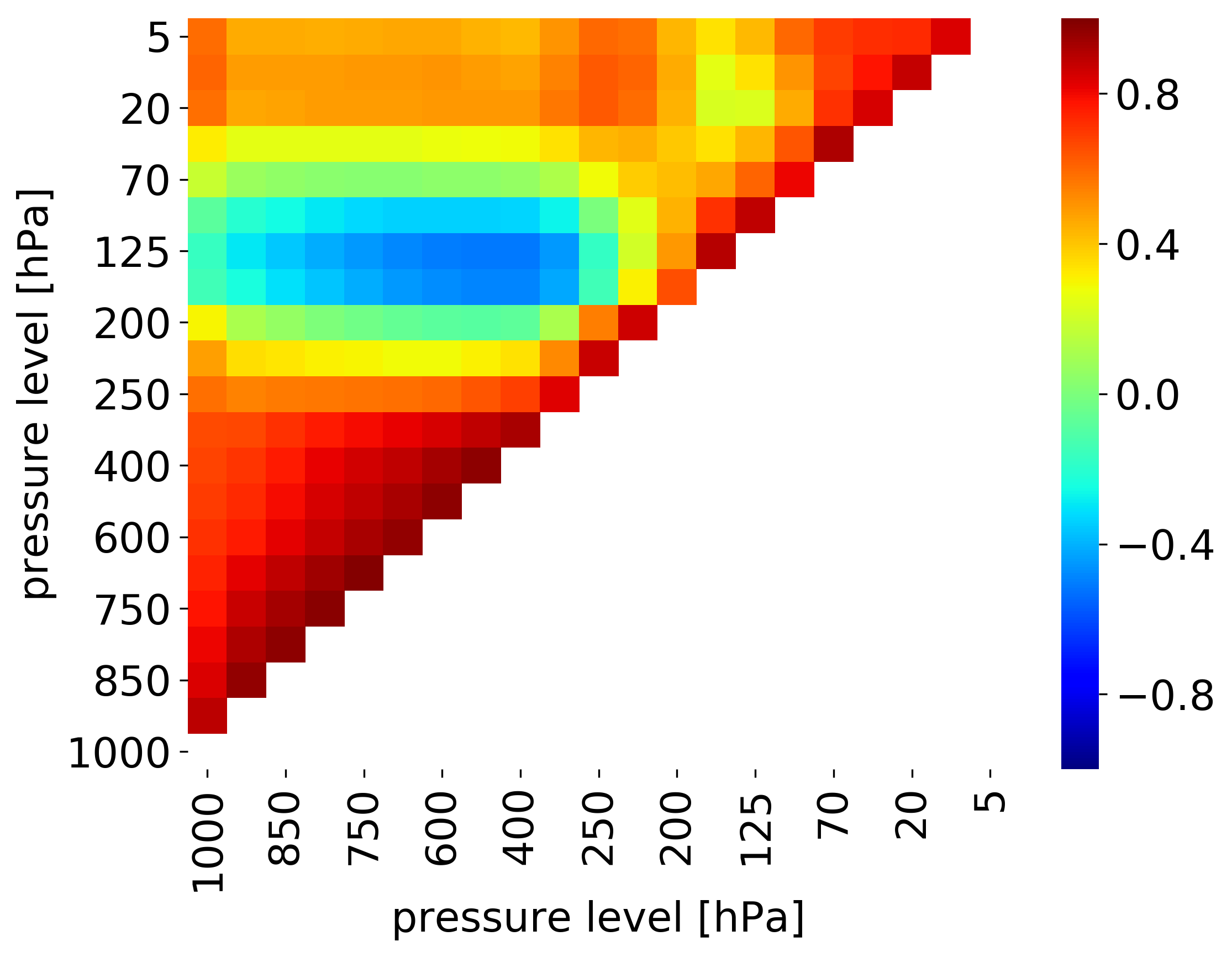}
\caption{Pairwise correlations between daily temperature variations $\Delta T$ of the different pressure levels considered in this analysis for Santiago de Compostela (from January 2015 to December 2016).}
\label{fig:layers}
\end{center}
\end{figure}

An experimental determination of $W_T(h)$ is not straightforward, and requires special statistical techniques, given the presence of strong correlations between the temperature variations of the different atmospheric layers. Figure \ref{fig:layers} shows the pairwise correlation matrix for the temperature variations of the different atmospheric layers on top of the detector. In general, the low stratosphere ($\sim$250-70 hPa) behaves opposite to the troposphere ($\sim$1000-250 hPa) and high stratosphere ($\sim$70-1 hPa). This is because the boundary layer ($\sim$925 hPa) is positively correlated with the rest of the troposphere through convection, while an increase in its temperature will generally result in the low stratosphere cooling down. This is a typical condition observed for latitude regions above 40$^\circ$ \cite{tropopause}.
Consequently, any attempt to obtain the temperature coefficients by means of a multivariate regression will result in coefficients whose values do not correspond to the actual values. If explanatory variables of a multiple regression model are strongly correlated, they provide redundant information and violate the condition of non-colinearity required in a least-squares regression. The coefficients will also be highly sensitive to small changes in the model and their sign will be dramatically dependent on the variables considered. In other words, slightly different models might lead to different conclusions. In this way, we would never know the actual effect of each variable. Clearly, any phenomenological model aimed at reliably describing the measured rates needs to start from a sensible set of uncorrelated temperature variables, that need to be obtained beforehand. We adapt for the task the Principal Components Regression (PCR) analysis, that has been successfully used before for this type of studies in \cite{yanchuk_pca, pca2019}.

\subsection{Barometric Effect}

Being much subtler, the temperature effect must be analyzed once the pressure effect has been removed. Moreover, in the case of gaseous detectors, the efficiency is a function of the ratio of electric field $E$ and pressure $P$ (represented by $E/P$ and dubbed reduced field), so even a high voltage and $T$-controlled environment is not sufficient to stabilize the detector response completely \cite{RPC}. The above dependency means that the detector efficiency is anticorrelated with pressure, and will add to the barometric effect at ground. Considering the atmospheric effect first, the relative change in the secondary c.r. rate caused by variations of the ground-level pressure has an exponential dependence. To first order approximation, it can be expressed through a linear relation:

\begin{equation}\label{eq:press_effect}
\frac{R}{R_0} = e^{\beta_{atm} \cdot \Delta P} \rightarrow  \frac{\Delta R}{R_0} \Big|_P \approx \beta \cdot \Delta P
\end{equation}
where $\frac{\Delta R}{R_0} \Big|_P$ is the relative variation of the c. r. rate due to the pressure effect, $R_0$ represents its average value over the period under consideration, $\Delta P=P-P_0$ is the deviation of the ground-level pressure with respect to its mean value ($P_0$) over the same period, and $\beta =\beta_{atm}+\beta_{det}$ is the barometric coefficient, with $\beta_{atm}$ representing the atmospheric effect and $\beta_{det}$ the detector contribution.

\begin{figure}[h!!!]
\begin{center}
\includegraphics[scale=0.45]{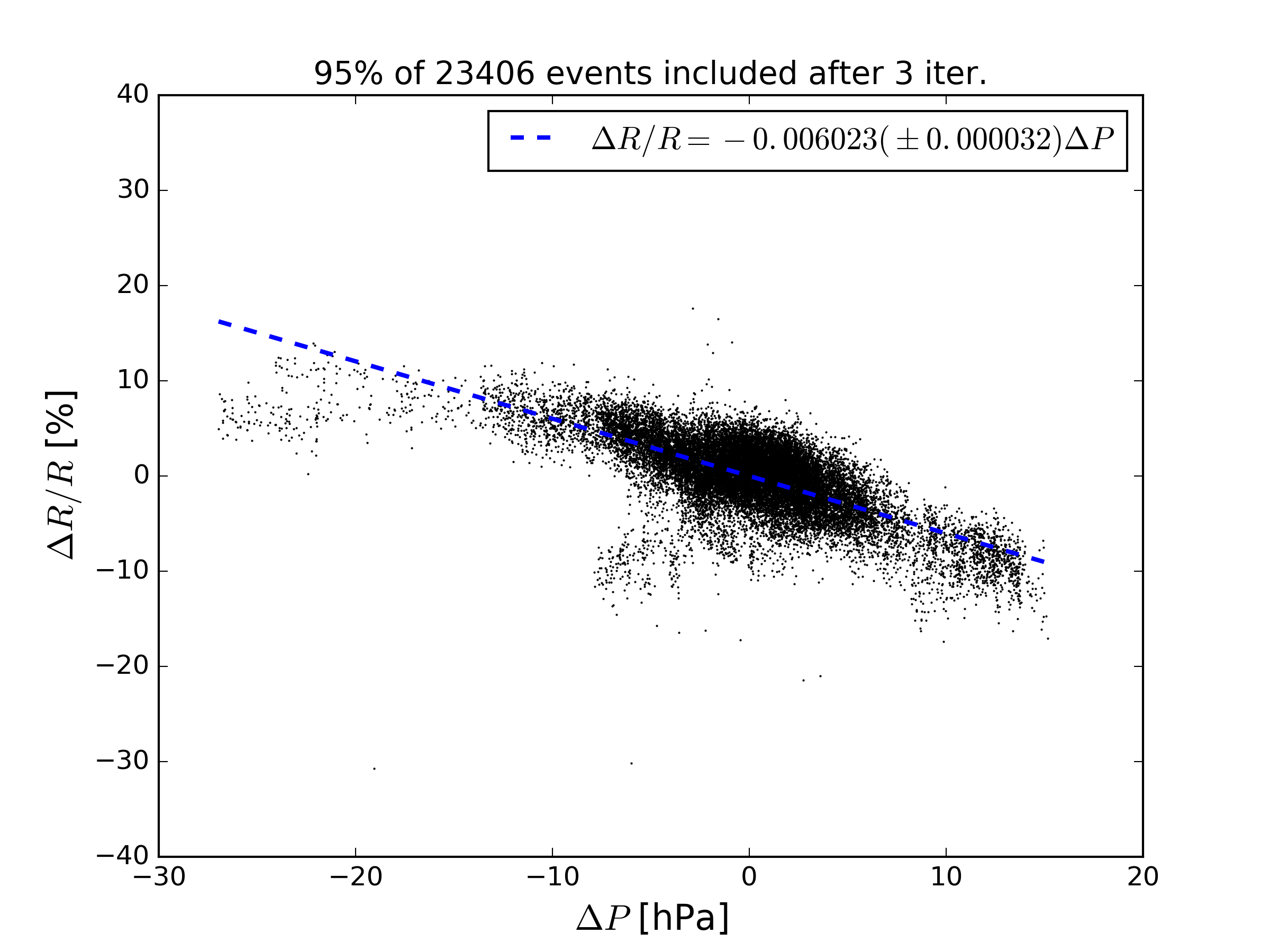}
\caption{Example of the linear fit method used to obtain the barometric coefficient for one of the sub-periods. The green lines delimit the points left out of a 2-$\sigma$ interval after an iterative procedure. The blue line is the resulting regression line.}
\label{fig:beta}
\end{center}
\end{figure}

The barometric coefficient was obtained separately for five different sub-periods, that displayed slightly different stability conditions. An iterative linear fit was performed, with data outside a 2-$\sigma$ interval removed from the fit  (Figure \ref{fig:beta}). Compatible barometric coefficients were obtained, whose mean value was determined to be $\beta= -0.59 \pm 0.02$ \%/hPa. This methodology allow us to remove any outliers in the data caused by detector instabilities and occasional space weather effects such as Forbush decreases or interplanetary events.

Finally, the barometric effect is removed using:
\begin{equation}
    \frac{\Delta R}{R_0} \Big|_T = \frac{\Delta R}{R_0} \Big|_{obs} - \frac{\Delta R}{R_0} \Big|_P
\end{equation}
where $\frac{\Delta R}{R_0} \Big|_{obs}$ are the experimental c.r. variations and  $\frac{\Delta R}{R_0} \Big|_T$ the remaining variations due to the temperature effect.

\subsection{Temperature effect} 

Variations of the measured rate of the secondary c.r. component due to the atmospheric temperature effect can be approximated by a linear combination of some temperature coefficients and the temperature variations of \textit{n} atmospheric layers \cite{dmitrievacoef}:

\begin{equation}\label{eq:model}
\frac{\Delta R}{R_0} \Big|_T = \sum_{i=1}^{n} W_T(h_i)\Delta T_i \Delta h_i
\end{equation}
where $\frac{\Delta R}{R_0} \Big|_T$  are the relative variations due to the temperature effect;  $W_T$, given in \% K$^{-1}$ atm$^{-1}$, is the corresponding temperature coefficient for the atmospheric layer \textit{i} at pressure $h_i$; $\Delta T_i=T_i-T_{0_i}$ are the temperature variations within the same layer with respect to its mean value ($T_{0_i}$), and $\Delta h_i=h_{i-1}-h_i$ is the layer thickness, in atm. 

Defining $k_{x_i}=W_T(h_i)\Delta h_i$, equation \ref{eq:model} can be rewritten as
\begin{equation}\label{eq:model2}
\frac{\Delta R}{R_0} \Big|_T = \sum_{i=1}^{n} k_{x_i}\Delta T_i 
\end{equation}
that we denote formally as
\begin{equation}\label{eq:model_matrix}
\mathbf{y}=\mathbf{X}\mathbf{k_x}
\end{equation}
where \textbf{y} is the vector of the measured relative variations $\frac{\Delta R}{R_0} \Big|_T$; \textbf{X} is the ($m \times n$) data matrix of the temperature variations whose columns are the temperature variations of the $ith$ pressure level and $\mathbf{k_x}$ refers to the vector of temperature coefficients, that we want to estimate. 

As mentioned earlier, the coefficients of this model can not be obtained by ordinary regression. For our purpose, we decided to use the Principal Component Regression (PCR) technique \cite{pca_book}. This method is applied when a dataset of variables shows multicollinearity, in our case the temperature variations. The idea is to build new uncorrelated variables (called principal components), maintaining the information conveyed by the original ones, and use them as the new predictors to estimate the unkown regression coefficients of the model. 

The PCA consists of an orthogonal linear transformation that converts the original variables to a new coordinate system. The principal components (PCs) represent the directions of the data containing the greatest variance. So, the first step is standardizing the $\Delta T_i$ measurements in \textbf{X}, dividing them by their standard deviations (over the analyzed period). This standardization is needed to prevent the variables with the highest variance from dominating. It causes a change in the notation, too. Too keep it simple, we mantain the current notation but taking into account that all the following calculations are based on standardized variables. 

The principal components are the eigenvectors (directions) obtained from the covariance matrix of \textbf{X} and sorted by the amount of explained variance. This set of orthogonal vectors forms a new basis in the new coordinate system. The matrix \textbf{X} can be transformed using the matrix of eigenvectors, defined as \textbf{A} ($n\times n$), in the following way

\begin{equation}\label{eq:pca_transf}
\mathbf{P=XA}
\end{equation}
where \textbf{P} is now the matrix ($m \times n$) containing the new variables in the new space. We got a set of uncorrelated variables because they were built using orthogonal eigenvectors. As a consequence, a new model can be built using variables \textbf{P}:
\begin{equation}\label{eq:model_pca}
\mathbf{y=Pk_p}
\end{equation}

Now, the new set of coefficients $\mathbf{k_p}$ can be obtained directly using a least-squares regression. Taking into account equation \ref{eq:pca_transf}, we can write

\begin{equation}\label{eq:model_pca2}
\mathbf{y=XAk_p}
\end{equation}

The regression coefficients $\mathbf{k_p}$ can be transformed back into the original space using equation \ref{eq:model_matrix} and \ref{eq:model_pca2}
\begin{equation}\label{eq:rel_coef}
\mathbf{k_x=Ak_p}
\end{equation}
and multiplying by standard deviations in order to go back to the original scale.

\begin{figure}
\begin{center}
\includegraphics[scale=0.55]{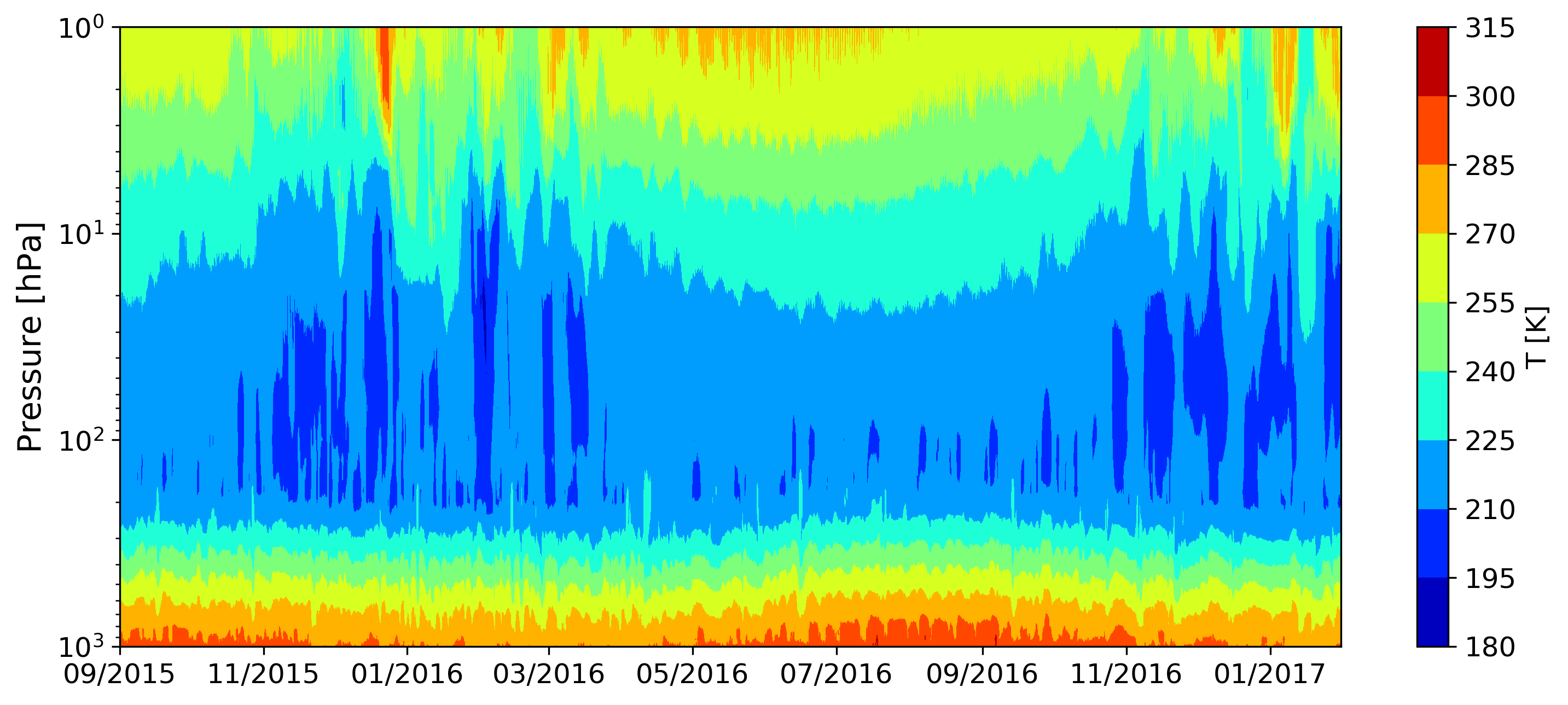}
\caption{Temperature time series of the atmosphere in Santiago de Compostela from October 2015 to January 2017.}
\label{fig:temp_map}
\end{center}
\end{figure}

\begin{figure}
\begin{center}
\includegraphics[scale=0.5]{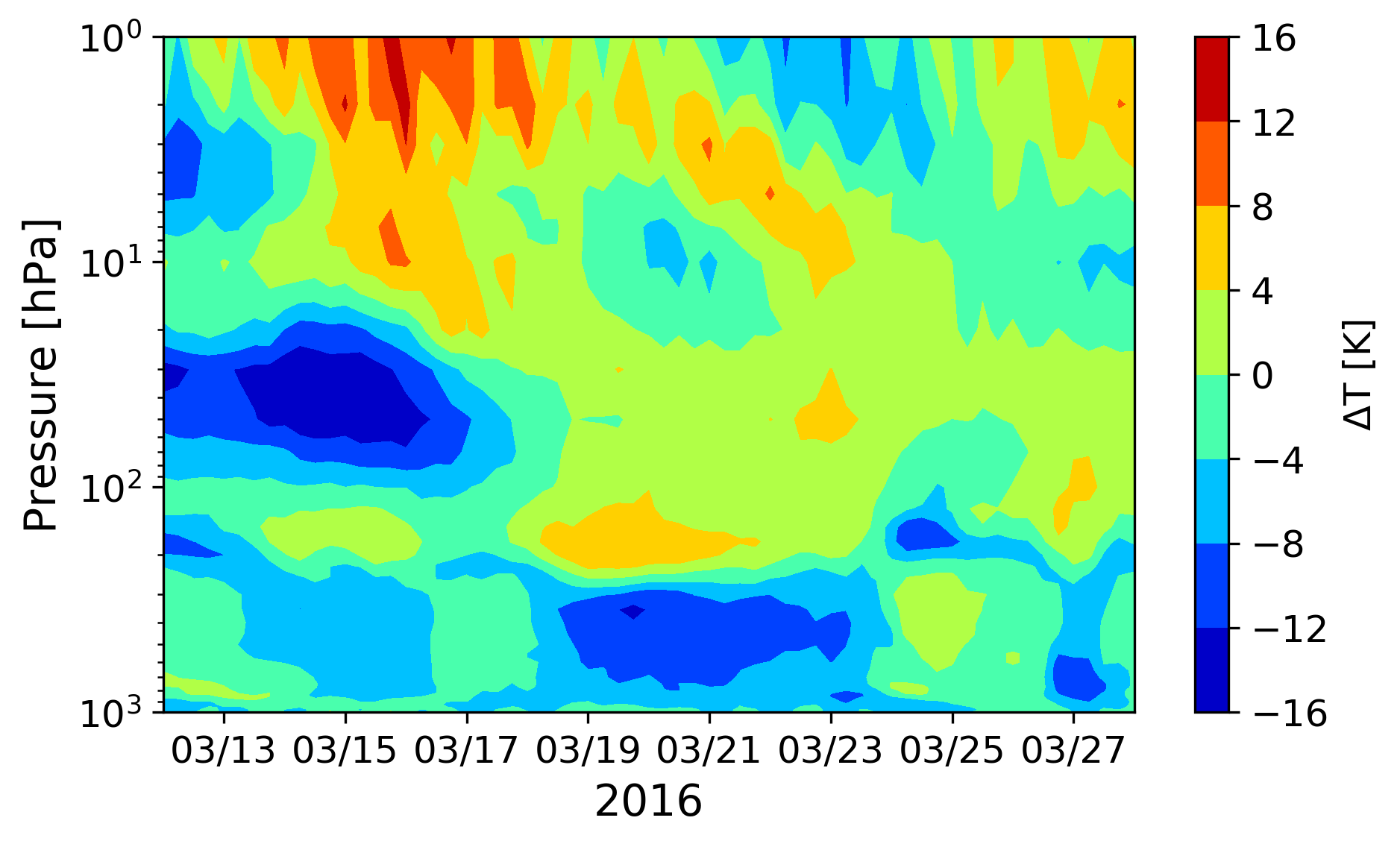}
\caption{Temperature anomaly for March 2016 were an example of a Sudden Stratospheric Warming is observed in the first half of the month. The upper stratosphere warms rapidly in a few days propagating way down into the troposphere in the next weeks.}
\label{fig:ssw}
\end{center}
\end{figure}

The year-to-year variability of the temperature data may affect the determination of the principal components, particularly if exceptional temperature changes took place during the data acquisition period, such as Stratospheric Warmings. This can be seen in Figures \ref{fig:temp_map} and \ref{fig:ssw}, where several stratospheric temperature anomalies (warmings and coolings) cab be seen during winter periods. Therefore, as a first step we use a training dataset from a time series of the last 30 years to determine the PCs of the temperature data and avoid the influence of outliers corresponding to exceptional events. 


PCR typically uses only a significant subset of all the principal components \textbf{P}$'$ to increase reliability. The components with higher variances are usually selected as the regressor variables for being the most important. No standard method exists for deciding how many components to retain. Anyhow, a good number of components should carry a high percentage of the total variance ($>$ 70\%). In order to decide the number of PCs to keep, we previously performed an analysis with reconstructed cosmic ray variations using a theoretical distribution of the temperature coefficients as a proxy \cite{dmitrievacoef}. These variations represent ideal data (i.e. without noise) only affected by the atmospheric temperature. Then, we apply the PCR method to these data to see how many PCs need to be kept in order to retrieve the original coefficients. To make this study more realistic, we follow the typical procedure of adding extra noise to the original data in three levels (low, medium and high) to be able to analyse the performance of the technique. It was observed that with two components it is possible to restore the correct values of the coefficients until an acceptable level of noise. Including more components destabilizes the result. 

Finally, the vector of coefficients $\mathbf{k'_p}$ is estimated by regressing the observed vector of cosmic ray data on the selected principal components $\mathbf{P'}$ using least-squares regression. So equation \ref{eq:model_pca} is reduced to

\begin{equation}
\mathbf{y=P'k'_p}
\end{equation}
where $\mathbf{P'}$ is now a matrix ($m \times r$) whose columns are the corresponding subset of columns of \textbf{P} (and $r<n$).

Using equation \ref{eq:rel_coef}, $\mathbf{k'_p}$ can be transformed back to the space of the actual temperature variables, providing the regression coefficients $\mathbf{k_x}$ that characterize the original model. Also, the relation $k_{x_i}=W_T(h_i)\Delta h_i$ introduced before is taken into account when converting the estimated regression coefficients to the distribution of temperature coefficients $W_T$, having a dimension \%/K$\cdot$atm. 

It must be noted that Partial Least Squares (PLS) regression could be an alternative to this technique because it is similar to PCR in that both select components that explain the most variance in the model. The difference is that PLS incorporates the response variable (the c.r. rate, in this case) into the analysis. One of the main reasons for not using this method is that our set of cosmic rays measurements is limited to a period of just two years. 

\section{Results and Discussion}

\begin{figure}
\includegraphics[scale=0.5]{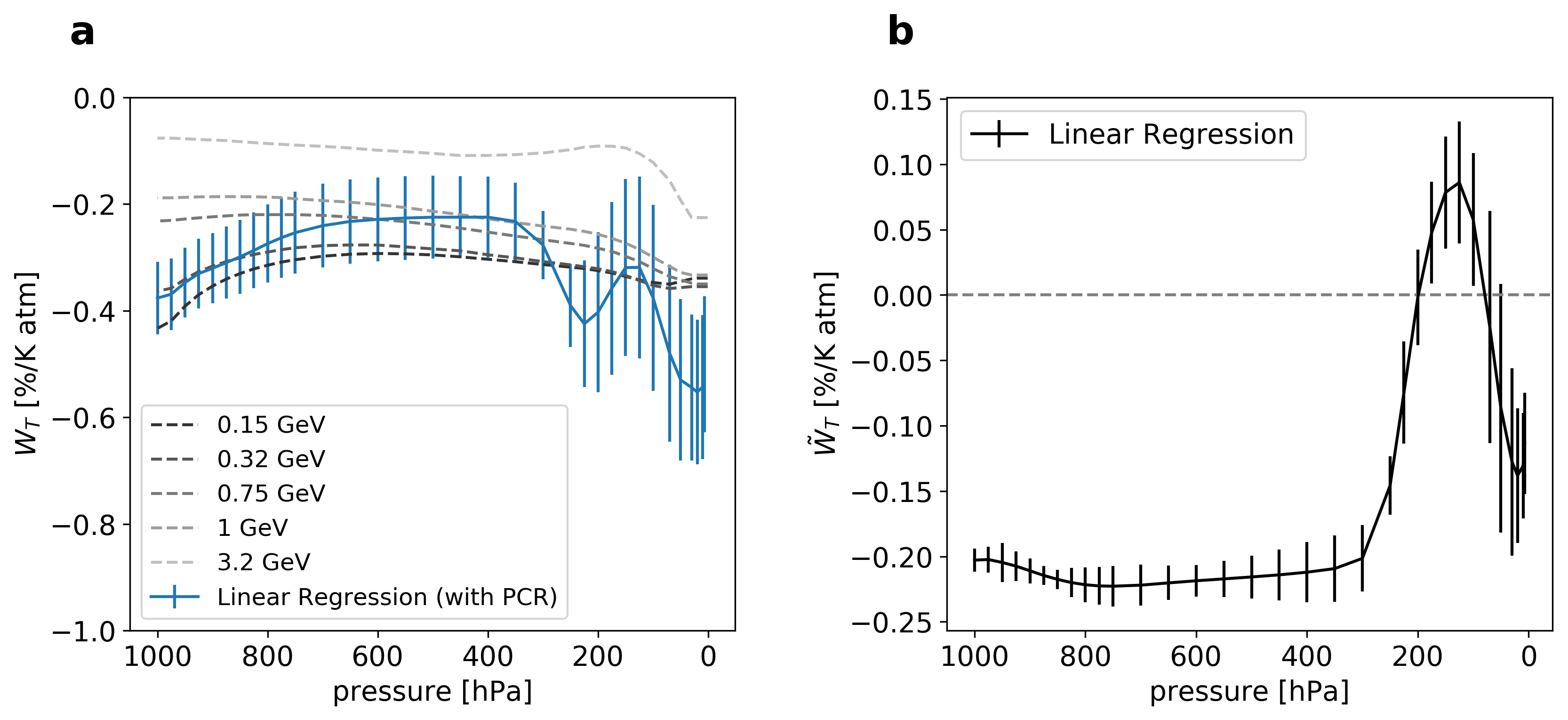}
\caption{(a) Distribution of temperature coefficients obtained with PCR for vertical tracks and comparison with the theoretical distribution for different energy thresholds: 0.15, 0.32, 0.75, 1 and 3.2 GeV. (b) Slopes $\tilde{W}_T$ obtained through a direct linear regression for the same data sample ($\theta<13^\circ$). }
\label{fig:coef&slopes}
\end{figure}

Figure \ref{fig:coef&slopes}a shows the distribution of temperature coefficients for the secondary c.r. component recorded at sea level for vertical incidence (blue line), where only events with a zenith angle $\theta$ lower than 13$^\circ$ were selected. The PCR method was applied to five different sub-periods of the data in order to account for systematic effects, expected to be mostly of instrumental origin at this stage, but also any remaining space weather phenomena having similar timescales to temperature variations. The average value and error bars are obtained from this combined analysis. 

As mentioned, the detector is placed in the first floor of a two-floor building. Therefore, the composition of the overburden material has been taken into account to estimate the value of the threshold energy, $E_{th} \sim$ 0.15 GeV. The theoretical distributions for different energy thresholds and zenith angle $\theta=0^\circ$ (grey lines in Figure \ref{fig:coef&slopes}a) as given in \cite{dmitrievacoef} are shown. A good agreement is observed for thresholds below 0.32 GeV, while above 0.75 GeV a systematic deviation appears, specially close to ground level.

Although being compatible, the estimated values in the troposphere ($>$300 hPa) are systematically above the theoretical ones. This might be a consequence of the method itself but could as well reflect the presence of the soft component in the measured rates, given that it anticipates a positive correlation with the lower layers of the atmosphere \cite{dorman}.  The values also differ at high altitudes, in this case due to the constraint in the selection of the number of principal components in the analysis. The PCR method computes the principal components taking into account the variance in the temperature data. Then, the first components reproduce the general variations in the troposphere and stratosphere (seasonal changes). Variations in the high atmosphere are considerably more complex than in the surface, as illustrated in Figure \ref{fig:layers}. Increasing the selected number of components would help to reduce this effect in an ideal situation. However, the optimum number of selected components in our case is the one that allows to obtain the best results of the coefficients without the solution being destabilized by noise and other instrumental effects. On the other hand, it should be noted that the accuracy of the ECMWF reanalysis is worse at high altitudes (0-200 hPa) due to the lack of data (less satellite/balloon observations, etc), so the temperatures have an inherent source of error that surely increases the difficulties in obtaining the coefficients at those heights. 

Figure \ref{fig:coef&slopes}b shows for illustration the slopes determined before applying the PCR. Each coefficient $\tilde{W}_T$ is obtained by direct regression between the relative variations of the c.r. intensity and the temperature variations for different layers:

\begin{equation}\label{eq:slopes}
\frac{\Delta R}{R_0} \Big|_T = \tilde{W}_T(h_i) \Delta T_i \Delta h_i 
\end{equation}
These coefficients are dominated by the multiple correlations between the atmospheric layers. In particular, the slopes in the troposphere are negative, become positive in the low stratosphere and return to negative values in the high stratosphere. The distribution of temperature coefficients, $W_T(h)$, indicates that the contribution of the troposphere to the total rate variation is higher than the rest ($\sum_{i=1}^{trop} \Delta h_i / \sum_{i=1}^n \Delta h_i \sim 75 \%$)  and the values of the coefficients for this layer are negative. Therefore, when a direct regression is performed, the slopes $\tilde{W}_T$ in the troposphere will be negative with values close to the real ones, $W_T$. The observed slopes in the high stratosphere will be negative due to the positive correlation with the troposphere (Figure \ref{fig:layers}). And the slopes in the low stratosphere are slightly positive due to the anticorrelation with the troposphere. A comparison with Figure \ref{fig:coef&slopes}b illustrates that these correlations have been largely removed with the PCR method.

Once the distribution of temperature coefficients $W_T$ has been calculated, we can study the temperature effect. For this purpose, equation \ref{eq:model} can be used to build the so-called ``effective temperature'', which allows to consider the entire atmospheric temperature profile through an unique parameter:

\begin{eqnarray}
\frac{\Delta R}{R_0} \Big|_T & = & \sum_{i=1}^{n} W_T(h_i) \Delta h_i \cdot \frac{\sum_{i=1}^n W_T(h_i)\Delta T_i \Delta h_i}{\sum_{i=1}^n W_T(h_i) \Delta h_i} = \\ \nonumber
& = & \alpha_T \Delta T_{eff}
\end{eqnarray}\label{eq:eq_teff}
Here the temperature coefficient $\alpha_T$ and the effective temperature $T_{eff}$ are defined as

\begin{equation}\label{eq:eff_coef}
\alpha_T = \sum_{i=1}^n W_T(h_i) \Delta h_i
\end{equation}

\begin{equation}\label{eq:teff}
T_{eff}=\frac{\sum_{i=1}^n W_T(h_i)T_i \Delta h_i}{\sum_{i=1}^n W_T(h_i) \Delta h_i}
\end{equation}
The definition of the effective temperature makes it possible to approximate the atmosphere as an isothermal body with a temperature $T_{eff}$, which is nothing but an average weighted by the product between the temperature coefficients and the atmospheric depth. It can be calculated using equation \ref{eq:teff} with the corresponding temperature coefficients $W_T$. Moreover, the theoretical value $\alpha_{T_{theor}}=-0.319$ \%/K is obtained using equation \ref{eq:eff_coef} for $E_{th}=0.15$ GeV. In our case, it can be experimentally calculated using the values of $W_T$ obtained by the PCR method: 
\[\alpha_{T_{exp}}=-0.279 \pm 0.051 \; \mathrm{\%/K}\]  
That is compatible with the theoretical one and also hints at the presence of the soft component, which could be the reason for the slight increase. 

\begin{table}[]
\caption{Values of the temperature coefficients $\alpha_T$ and $\alpha_{MSS}$ obtained in this work, together with the deviation from independent estimates, dubbed $\Delta \alpha$.}
\centering
\begin{tabular}{ccc|c}
\hline
 & Theoretical \cite{dmitrievacoef} & This work & $\Delta \alpha$  \\ 
$\alpha_T$ (\%/K)  & -0.319  & -0.279 $\pm$ 0.051 & 0.040 $\pm$ 0.051 \\ 
& & & \\ 
 & \cite{mendonca_rigidity} & This work & $\Delta \alpha$  \\ 
$\alpha_{MSS}$ (\%/K) & -0.271 &  -0.233 $\pm$ 0.045 & 0.038 $\pm$ 0.045 \\ \hline
\end{tabular}
\label{tab:alphas}
\end{table}

A complementary approach exists using the so-called mass-weighted temperature $T_{MSS}$ and its corresponding coefficient $\alpha_{MSS}$ \cite{mendonca}. In this case, the c.r. variations due to the temperature effect are approximated by

\begin{equation}\label{eq:eq_tmss}
\frac{\Delta R}{R_0} \Big|_T = \alpha_{MSS} \Delta T_{MSS}
\end{equation}

\begin{equation}
T_{MSS} = \sum_{i=1}^n T_i \cdot \left( \frac{x(h_i)-x(h_{i+1})}{x(h_0)} \right) 
\end{equation}
where $x(h_i)$ is the atmospheric depth at the same altitude. Indeed, a very recent study from the GMDN \cite{mendonca_rigidity} found a relation between the mass-weighted temperature coefficient and the cutoff rigidity ($R_c$) and latitude ($L$) for vertical incidence:

\begin{equation}\label{eq:coeffmss}
\alpha_{MSS} = -0.304 + 0.0389 \cdot \ln{R_c}-0.0488 \cdot \sin{L}    
\end{equation}

Introducing in this equation the values for $R_c$ and $L$ corresponding to the location of our detector, we anticipate a temperature coefficient of -0.271 \%/K. On the other hand, application of a linear regression to our data following equation \ref{eq:eq_tmss} gives $\alpha_{MSS}=-0.233 \pm 0.045$ \%/K, which again points to contamination from the soft component. Our analysis is summarised in Table \ref{tab:alphas}, highlighting the deviation from the other two estimates discussed in this work ($\Delta \alpha$). The fact that $\Delta \alpha$ is similar for both $\alpha_{MSS}$ and $\alpha_T$, suggests a common origin to the observed excess.

\begin{figure}
\includegraphics[width=\textwidth]{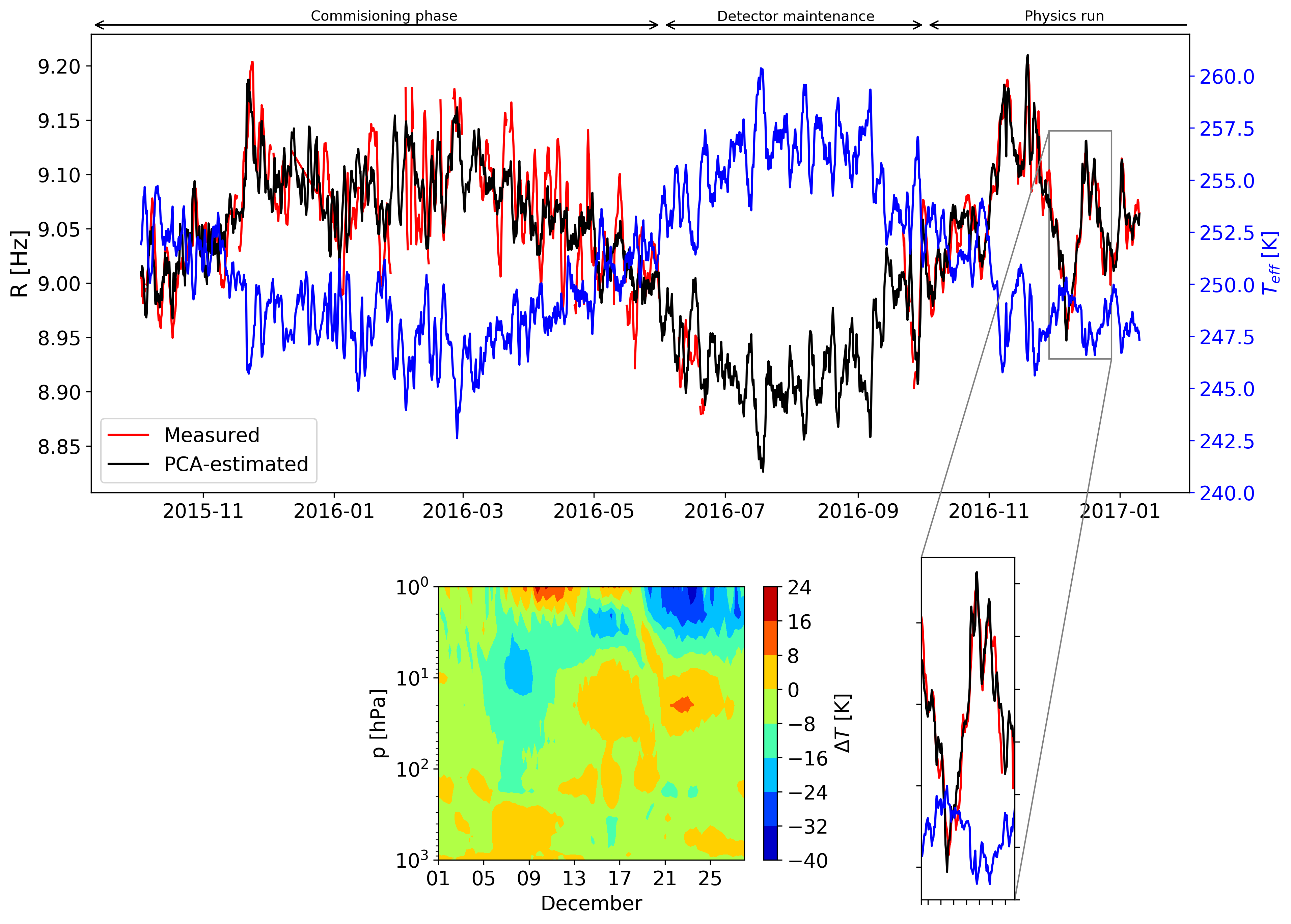}
\caption{Top: cosmic ray rate corrected by pressure as observed by the TRAGALDABAS detector (red curve) and calculated via PCR method (black curve) compared with the effective temperature (blue curve). Bottom: zoom into the December 2016 Sudden Warming event, and P, $\Delta$T map on that period.}
\label{fig:data}
\end{figure}

Figure \ref{fig:data} shows the secondary c.r. after pressure correction compared with the estimated temperature effect modeled through our two main PCs and their corresponding regression coefficients. The evolution of the effective temperature is shown as well. It is possible to appreciate the typical seasonal behavior reported by other detectors: c.r. rate reaches its maximum in winter when the atmosphere is colder while it declines towards summer when it is warmer. Moreover, the procedure is able to interpolate the temperature effect in those periods of missing data, such us July and August of 2016. 
Overall, the estimated temperature effect is able to reproduce $\sim77$\% of the variability of the observed data, demonstrating that the PCR is a reasonable method to recover the seasonal variability.

A significant behaviour is also observed at the end of 2016: rates show an abrupt decrease followed by a major increase (see zoom in Figure \ref{fig:data}). The PCR method is able to reproduce this trend as well. If the temperature variations of this period are analysed, several cooling and warmings are observed: a Sudden Stratospheric Warming in the high stratosphere takes place, followed by a great cooling of at least -40 K. However, an opposite behaviour is observed in the low stratosphere. 

The sign of the calculated distribution of the coefficients $W_T$ (Figure \ref{fig:coef&slopes}a) indicates that the negative temperature effect dominates throughout the atmosphere. This means that, if the temperature varies in any atmospheric layer, the measured rate will vary in inverse proportionallity to that indicated by the corresponding coefficient. Bearing this in mind, we can expect a measured rate decrease due to the increase of temperature in the troposphere and high stratosphere. However, the reduction of the temperature of the low stratosphere, given the presence of correlations between layers, will tend to increase the rates. The effective temperature describes the global effect of these different temperature variations. In this case, as the layers of the troposphere have more weight ($\sim$75\% of the total atmospheric mass) with respect to the stratosphere, their effect is dominant and that is why the seasonal evolution of the effective temperature in Figure \ref{fig:data} is similar to the surface temperature, and appears anticorrelated with the observed rates.

The distribution of temperature coefficients can be used, along with the barometric coefficient, to remove the atmospheric effects and analyse space weather phenomena. This is briefly described in \ref{sec:appen} with the analysis of a Forbush Decrease (FD) and will be the subject of future works.

To sum up, this is the first demonstration that the timing RPC technology can be succesfully applied to studies about the atmosphere condition, in particular its temperature profile. With the commissioning phase already over and the detector fully operative, additional work could focus on carrying out differential studies on the angular response as well as carefully correcting for space weather phenomena.

\section{Conclusions}

We have commissioned and calibrated a small-size 2 m$^2$ timing RPC detector devoted to the detailed study of cosmic rays at ground level, and performed the first analysis of the atmospheric temperature effect with this technology. By studying a data sample of barely one year, it has been possible to estimate the distribution of temperature coefficients ($W_T(h)$), showing that the contribution of the hard component is dominant and in good agreement with theoretical calculations.

We show how the presence of strong correlations among the different atmospheric layers precludes the use of conventional regression methods. A Principal Component Regression (PCR), considering the first two components, is sufficient to capture at least 77\% of the variability, giving a good description of the $W_T(h)$ and the global slope parameter $\alpha_{T_{exp}}=-0.279 \pm 0.051$ \%/K (compared to a theoretical value of $\alpha_{T_{theor}}=-0.319$ \%/K). This results in an anticorrelation with the effective atmospheric temperature, that allows to clearly identify its seasonal cycles as well as short-term exceptional events (such as the tropospheric consequences of a Sudden Stratospheric Warming), through measurements performed at ground level. 

\appendix
\section{Monitoring a Forbush Decrease}\label{sec:appen}

As mentioned in the introduction, c.r. measurements provide valuable data for different research areas, in particular space weather. But as we have seen so far, when analyzing variations in c.r. intensity using ground-based detectors, atmospheric effects cannot be ignored. The pressure and temperature effects produce significant variations. Therefore, it is important to remove those in order to study any solar or interplanetary phenomena \cite{dorman}. 

In June 2015 the Sun was very active and produced a significant number of coronal mass ejections towards the Earth, initiating a large Forbush Decrease (FD) event (more information at \cite{soho_nasa}). A FD may be caused when a solar disturbance travels away from the Sun towards the Earth, affecting the galactic cosmic ray flux, which conveys the most energetic particles coming from outside the heliosphere. Such disturbance will produce a region of suppressed c.r. density located downstream of the coronal mass ejection, behind the interplanetary shock which this fast ejection produces in the medium ahead of it. In such a case, the c.r. intensity at ground shows a fast decrease, reaching a minimum within about a day, followed by a slow recovery phase lasting for several days \cite{forbush}.

\begin{figure}
    \centering
    \includegraphics[scale=0.55]{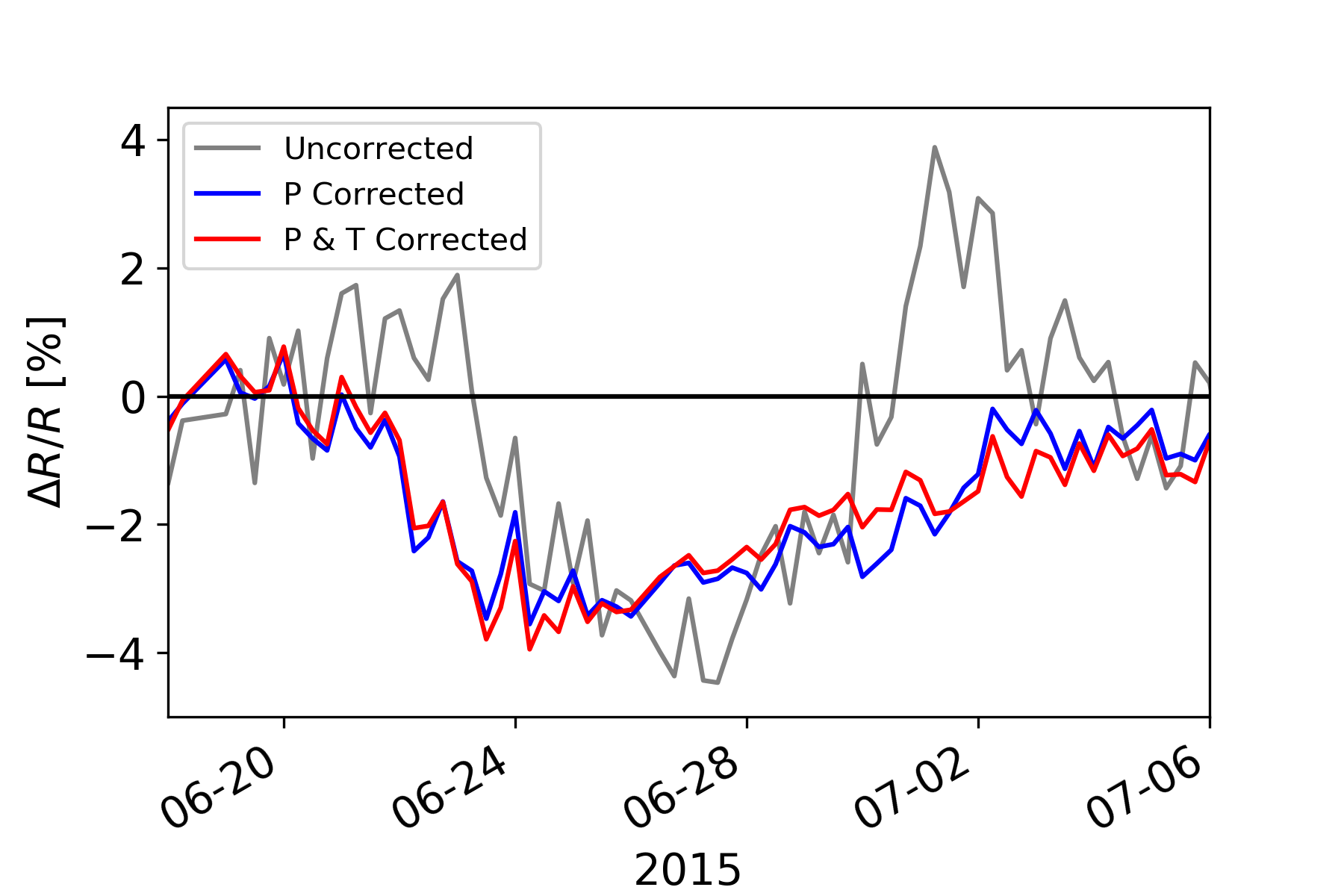}
    \caption{Forbush Decrease event on 22 June 2015: uncorrected (grey curve), only pressure corrected (blue curve) and pressure and temperature corrected variations (red curve).}
    \label{fig:forbush}
\end{figure}

The decrease in the TRAGALDABAS counting rate on 22 June 2015 is the first FD registered over the period from 2015 to 2017. Figure \ref{fig:forbush} shows the relative variations before corrections (grey), corrected only by pressure (blue) and corrected by both temperature and pressure (red) in the period from 18 June to 6 July, 2015. A fast decreasing phase is observed after pressure corrections, reaching a minimum in a couple of days of about $\sim$4\%, and followed by a slow recovery phase during the next days. Without atmospheric corrections, this characteristic FD behaviour is not discernible.

\acknowledgments

We gratefully acknowledge The TRAGALDABAS Collaboration and J.A Garz\'on-Heydt (LabCAF-Universtiy of Santiago de Compostela, IGFAE) for providing the data for this work. \\
We thank financial support by the Spanish Ministerio de Econom\'ia y Competitividad and European Regional Development Fund under contract RTI2018-097063-B100 AEI/FEDER, UE, and by Xunta de Galicia under Research Grant No. 2018-PG082. I.R. and V.P.M. are part of the CRETUS Strategic Partnership (AGRUP2015/02). All these programs are co-funded by FEDER (UE).\\
D.G.C. thanks to the Ministerio de Ciencia, Investigaci\'on y Universidades and the European Social Fund (FSE) for a predoctoral grant (FPI 2017).\\
D.G.D. acknowledges the Ramon y Cajal program, contract RYC-2015-18820.\\
The data are in these repositories and cited in the references: 
the ECMWF ERA-Interim reanalysis data were obtained from \\
https://www.ecmwf.int/en/forecasts/datasets/reanalysis-datasets/era-interim; \\
Cosmic ray rates can be found at http://dx.doi.org/10.17632/f93rp64p6z.2.

%% ------------------------------------------------------------------------ %%
%% References and Citations

%%%%%%%%%%%%%%%%%%%%%%%%%%%%%%%%%%%%%%%%%%%%%%%
%
% \bibliography{<name of your .bib file>} don't specify the file extension
%
% don't specify bibliographystyle
%%%%%%%%%%%%%%%%%%%%%%%%%%%%%%%%%%%%%%%%%%%%%%%

\bibliography{bibliografia}

%Reference citation instructions and examples:
%
% Please use ONLY \cite and \citeA for reference citations.
% \cite for parenthetical references
% ...as shown in recent studies (Simpson et al., 2019)
% \citeA for in-text citations
% ...Simpson et al. (2019) have shown...
%
%
%...as shown by \citeA{jskilby}.
%...as shown by \citeA{lewin76}, \citeA{carson86}, \citeA{bartoldy02}, and \citeA{rinaldi03}.
%...has been shown \cite{jskilbye}.
%...has been shown \cite{lewin76,carson86,bartoldy02,rinaldi03}.
%... \cite <i.e.>[]{lewin76,carson86,bartoldy02,rinaldi03}.
%...has been shown by \cite <e.g.,>[and others]{lewin76}.
%
% apacite uses < > for prenotes and [ ] for postnotes
% DO NOT use other cite commands (e.g., \citet, \citep, \citeyear, \nocite, \citealp, etc.).
%

\end{document}

%% ------------------------------------------------------------------------ %%
%
%  TITLE
%
%% ------------------------------------------------------------------------ %%

%\includegraphics{agu_pubart-white_reduced.eps}

\title{Supporting Information for "Insert Title"}
%
% e.g., \title{Supporting Information for "Terrestrial ring current:
% Origin, formation, and decay $\alpha\beta\Gamma\Delta$"}
%
%DOI: 10.1002/%insert paper number here%

%% ------------------------------------------------------------------------ %%
%
%  AUTHORS AND AFFILIATIONS
%
%% ------------------------------------------------------------------------ %%

% List authors by first name or initial followed by last name and
% separated by commas. Use \affil{} to number affiliations, and
% \thanks{} for author notes.
% Additional author notes should be indicated with \thanks{} (for
% example, for current addresses).

% Example: \authors{A. B. Author\affil{1}\thanks{Current address, Antartica}, B. C. Author\affil{2,3}, and D. E.
% Author\affil{3,4}\thanks{Also funded by Monsanto.}}

\authors{=Authors=}

% \affiliation{1}{First Affiliation}
% \affiliation{2}{Second Affiliation}
% \affiliation{3}{Third Affiliation}
% \affiliation{4}{Fourth Affiliation}

\affiliation{=number=}{=Affiliation Address=}
%(repeat as many times as is necessary)

%% ------------------------------------------------------------------------ %%
%
%  BEGIN ARTICLE
%
%% ------------------------------------------------------------------------ %%

% The body of the article must start with a \begin{article} command
%
% \end{article} must follow the references section, before the figures
%  and tables.

\begin{article}

%% ------------------------------------------------------------------------ %%
%
%  TEXT
%
%% ------------------------------------------------------------------------ %%

\noindent\textbf{Contents of this file}
%%%Remove or add items as needed%%%
\begin{enumerate}
\item Text S1 to Sx
\item Figures S1 to Sx
\item Tables S1 to Sx
%if Tables are larger than 1 page, upload as separate excel file
\end{enumerate}
\noindent\textbf{Additional Supporting Information (Files uploaded separately)}
\begin{enumerate}
\item Captions for Datasets S1 to Sx
\item Captions for large Tables S1 to Sx (if larger than 1 page, upload as separate excel file)
\item Captions for Movies S1 to Sx
\item Captions for Audio S1 to Sx
\end{enumerate}

\noindent\textbf{Introduction}
%Type or paste your text here. The introduction gives a brief overview of the supporting information. You should include information %about as many of the following as possible (when appropriate):
% 1. a general overview of the kind of data files;
% 2. information about when and how the data were collected or created;
% 3. a general description of processing steps used;
% 4. any known imperfections or anomalies in the data.

%\clearpage

%Delete all unused file types below. Copy/paste for multiples of each file type as needed.
\noindent\textbf{Text S1.}
%Type or paste text here. This should be additional explanatory text, such as: extended descriptions of results, full details of models, extended lists of acknowledgements etc.  It should not be additional discussion, analysis, interpretation or critique. It should not be an additional scientific experiment or paper.
%
%Repeat for any additional Supporting Text

%%Enter Data Set, Movie, and Audio captions here
%%EXAMPLE CAPTIONS

\noindent\textbf{Data Set S1.} %Type or paste caption here.
%upload your dataset(s) to AGU's journal submission site and select "Supporting Information (SI)" as the file type. Following naming %convention: ds01.

%Repeat for any additional Supporting data sets

\noindent\textbf{Movie S1.} %Type or paste caption here.
%upload your movie(s) to AGU's journal submission site and select, "Supporting Information %(SI)" as the file type. Following naming convention: ms01.

%Repeat any additional Supporting movies

\noindent\textbf{Audio S1.} %Type or paste caption here.
\end{article}
\clearpage